\RequirePackage{fix-cm}
\documentclass[twocolumn]{svjour3}          

\smartqed  

\usepackage{graphicx}
\usepackage{latexsym}
\usepackage{amsmath}
\usepackage{amssymb}
\usepackage{multirow}

\journalname{Brazilian Journal of Physics}

\begin{document}

\title{Generating pseudo-random discrete probability distributions}
\subtitle{About the iid, normalization, and trigonometric methods}
\author{Jonas Maziero }

\institute{Jonas Maziero \at Departamento de F\'isica, Universidade Federal de Santa Maria, 97105-900, Santa Maria, RS, Brazil \\ \email{jonas.maziero@ufsm.br}}

\date{Received: date / Accepted: date}

\maketitle

\begin{abstract}
The generation of pseudo-random discrete probability distributions
is of paramount importance for a wide range of stochastic simulations
spanning from Monte Carlo methods to the random sampling of quantum
states for investigations in quantum information science. In spite of
its significance, a thorough exposition of such a procedure is lacking
in the literature. In this article we present relevant details concerning
the numerical implementation and applicability of what we call the iid, normalization, and
trigonometric methods for generating an unbiased probability vector $\mathbf{p}=(p_{1},\cdots,p_{d})$. An immediate application of these results regarding the generation of pseudo-random pure quantum states is also described.


\end{abstract}

\section{Introduction}

Roughly speaking, randomness is the fact that, even using all the information that we have about a physical system, in some situations it is impossible, or unfeasible, for us to predict exactly what will be the future state of that system. Randomness is a facet of nature that is ubiquitous and very influential
in our and other societies \cite{Debora_Randomness,Mlodinow_Randomness,Bell}.
As a consequence, it is also an essential aspect of our science and
technology. The related research theme, that was motivated initially
mainly by gambling and led eventually to probability theory \cite{DeGroot,Jaynes_Prob},
is nowadays a crucial part of many different fields of study such
as computational simulations, information theory, cryptography, statistical
estimation, system identification, and many others \cite{Binder_MC,Cover-Thomas,Devore_Prob_App,Ribeiro,Oliveira,Resende,Ellis}.

One rapid-growing area of research for which randomness is a key concept
is the maturing field of quantum information science (QIS). Our main
aim in this interdisciplinary field is understanding how quantum systems can be harnessed
in order to use all Nature's potentialities for information storage,
processing, transmission, and protection \cite{Nielsen_Book,Wilde_Book,Maziero_bjp1}. 

Quantum mechanics \cite{Peres,Sakurai} is one of the present fundamental
theories of Nature. The essential mathematical object in this theory
is the density operator (or density matrix) $\rho$. It embodies all
our knowledge about the preparation of the system, i.e., about its
state. 
From the mathematical
point of view, a density matrix is simply a positive semi-definite
matrix (notation: $\rho\ge0$) with trace equal to one ($\mathrm{Tr}(\rho)=1$).
Such kind of matrix can be written as $\rho={\textstyle \sum_{j}}r_{j}\Pi_{j}$, which is known as the spectral decomposition of $\rho$. In the last
equation $\Pi_{j}$ is the projector ($\Pi_{j}\Pi_{k}=\delta_{jk}\Pi_{j}$
and $\sum_{j}\Pi_{j}=\mathbb{I}_{d}$, where $\mathbb{I}_{d}$ is
the $d-$dimensional identity matrix) on the vector subspace corresponding
to the eigenvalue $r_{j}$ of $\rho$. From the positivity of $\rho$
follows that, besides it being Hermitian and hence having real eigenvalues,
its eigenvalues are also non-negative, $r_{j}\ge0$. Once the trace
function is base independent, the eigenvalues of $\rho$ must sum
up to one, $\mathrm{Tr}(\rho)=\textstyle{\sum_{j}}r_{j}=1$. Thus we see that
the set $\{r_{j}\}$ possesses all the features that define a probability
distribution (see e.g. Ref. \cite{DeGroot}).

The generation of pseudo-random quantum states is an essential tool
for inquires in QIS (see e.g. Refs. \cite{James_rand_r,Ramos_rand_r,Adesso_ObsQD,Adesso_Adisc,Plastino_EqEnt,Plastino_rbits,Plastino_MES,Genovese_rqs,Rau_rqs,Wang_rqs,Boyd,Englert1,Englert2})
and involves two parts. The first one is the generation of pseudo-random
sets of projectors $\{\Pi_{j}\}$, that can be cast in terms of the
creation of pseudo-random unitary matrices. There are several methods
for accomplishing this last task \cite{Stewart_rand_U,Zyczkowski_U,Petruccione_PofU,Lloyd_rand_U},
whose details shall not be discussed here. Here we will
address the second part, which is the generation of pseudo-random
discrete probability distributions \cite{Vedral_Prob_trigo,Zyczkowski_Vol1,Fritzche4},
dubbed here as pseudo-random probability vectors (pRPV).

In this article we go into
the details of three methods for generating numerically pRPV. We present the problem details in Sec. \ref{sec_problem}. The Sec. \ref{sec_iid} is devoted to present a simple method, the iid method, and to show that it is not useful for the task regarded in this article. In Sec.
\ref{sec_norm} the standard normalization method is discussed. The
bias of the pRPV appearing in its naive-direct implementation is highlighted.
A simple solution to this problem via random shuffling of the pRPV
components is then presented. In Sec. \ref{sec_trigo} we consider
the trigonometric method. After discussing some issues regarding its
biasing and numerical implementation, we study and compare the probability distribution generated and the computer time required by the last two methods when the dimension of the pRPV is
varied. The conclusions and prospects are presented in Sec. \ref{conclusions}.

\section{The problem}
\label{sec_problem}
By definition, a discrete probability distribution \cite{DeGroot}
is a set of non-negative real numbers, 
\begin{equation}
p_{j}\ge0,
\label{eq:prob1}
\end{equation}
that sum up to one,
\begin{equation}
{\textstyle \sum_{j=1}^{d}} p_{j} = 1. 
\label{eq:prob2}
\end{equation}
In this article we will utilize the numbers $p_{j}$ as the components
of a probability vector
\begin{equation}
\mathbf{p} = ( p_{1}, \cdots, p_{d} ).
\end{equation}

Despite the nonexistence of consensus regarding the meaning of probabilities
\cite{DeGroot}, here we can consider $p_{j}$ simply as the relative
frequency with which a particular value $x_{j}$ of a physical observable
modeled by a random variable $X$ is obtained in measurements of that
observable under appropriate conditions.

We would like to generate numerically a pseudo-random probability
vector $\mathbf{p}$ whose components $\{p_{j}\}_{j=1}^{d}$ form a probability
distribution, i.e., respect Eqs. (\ref{eq:prob1}) and (\ref{eq:prob2}).
In addition we would like the pRPV to be unbiased, i.e., the components of $\mathbf{p}$ must have similar probability distributions. A necessary condition for fulfilling this
last requisite is that the average value of $p_{j}$ (notation: $\langle p_{j}\rangle$)
becomes closer to $1/d$ as the number of pRPV generated becomes large.

At the outset we will need a pseudo-random number generator (pRNG).
In this article we use the Mersenne Twister pRNG \cite{Matsumoto_MT},
that yields pseudo-random numbers (pRN) with uniform distribution
in the interval $[0,1]$. We observe however that the results reported in article can be applied straightforwardly also when dealing with true random numbers \cite{Benson_QRNG,Miszczak_QRNG}.

\section{The iid method}
\label{sec_iid}

A simple way to generate an unbiased pseudo-random probability vector $\mathbf{p}=(p_{1},\cdots,p_{d})$ is as follows. If we create $d$ \emph{independent} pseudo-random numbers $x_{j}$ with \emph{identical} probability \emph{distributions} in the interval $[0,1]$ (so the name of the method) and set
\begin{equation}
p_{j} := \frac{x_{j}}{\textstyle{\sum_{k=1}^{d}x_{k}}},
\end{equation}
we will obtain a well defined discrete probability distribution, i.e., $p_{j}\ge 0$ and $\sum_{j=1}^{d}p_{j}=1$. Besides, as $\mathbf{p}$ is unbiased, the mean value of $p_{j}$ approaches $1/d$ as the number of samples grows.

\begin{figure}[b]
\centering
\includegraphics[scale=0.37]{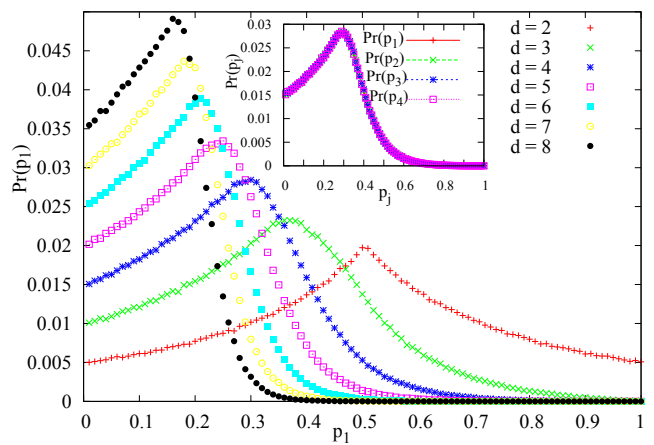}
\protect\caption{(color online) Probability distribution for the first component of the unbiased probability vector $\mathbf{p}=(p_{1},\cdots,p_{d})$ for one million random samples generated using the iid method. In the inset is shown the probability distribution for the components of the probability vector $\mathbf{p}=(p_{1},p_{2},p_{3},p_{4})$.}
\label{fig_stat_prob_dist_iid}
\end{figure}

Nevertheless, we should note that the sum $\sum_{k=1}^{d}x_{k}$ shall be typically greater than one. This in turn will lead to the impossibility for the occurrence of pRPVs with one of its components equal (or even close) to one. As can be seem in Fig. \ref{fig_stat_prob_dist_iid}, this problem becomes more and more important as $d$ increases. Therefore, this kind of drawback totally precludes the applicability of the iid method for the task regarded in this article.

\section{The normalization method}

\label{sec_norm} 

Lets us begin our discussion of this method by considering a probability vector with dimension $d=2$, i.e., $\mathbf{p}=(p_{1},p_{2})$. If the pRNG
is used to obtain $p_{1}\in[0,1]$ and we impose the normalization
to get $p_{2}=1-p_{1}$, we are guaranteed to generate an uniform
distribution for $p_{j}\in[0,1]$ for both $j=1$ and $j=2$. 

If $d=3$ then $\mathbf{p}=(p_{1},p_{2},p_{3})$ and the pRNG is used
again (two times) to obtain $p_{1}\in[0,1]$ and $p_{2}\in[0,1-p_{1}]$.
Note that the interval for $p_{2}$ was changed because of the \emph{normalization}
of the probability distribution, which is also used to write $p_{3}=1-(p_{1}+p_{2})$.
As $p_{1}$ is equiprobable in $[0,1]$, for a large number of samples
of the pRPV, its mean value will be $1/2$. This shall restrict the
values of the other components of $\mathbf{p}$, shifting the ``center''
of their probability distributions to $1/4$, biasing thus the pRPV.
Of course, if one increases the dimension of the pRPV, the same effect
continues to be observed, as is illustrated in the table below for
$10^{6}$ pRPV generated for each value of $d$.

\begin{center}
\begin{tabular}{|c|c|c|c|c|c|c|}
\hline 
$d$ & $\langle p_{1}\rangle$ & $\langle p_{2}\rangle$ & $\langle p_{3}\rangle$ & $\langle p_{4}\rangle$ & $\langle p_{5}\rangle$ 
\tabularnewline
\hline 
\hline 
$2$ & $0.5001$ & $0.4999$ &  & \multirow{2}*{} & \multirow{3}*{} 
\tabularnewline
\cline{1-4} 
$3$ & $0.5000$ & $0.2499$ & $0.2501$ &  &   \tabularnewline
\cline{1-5} 
$4$ & $0.4999$ & $0.2503$ & $0.1248$ & $0.1250$ &   \tabularnewline
\cline{1-6} 
$5$ & $0.4998$ & $0.2501$ & $0.1252$ & $0.0625$ & $0.0624$  \tabularnewline
\hline 
\end{tabular}
\par\end{center}

\begin{figure}[b]
\begin{center}
\includegraphics[scale=0.36]{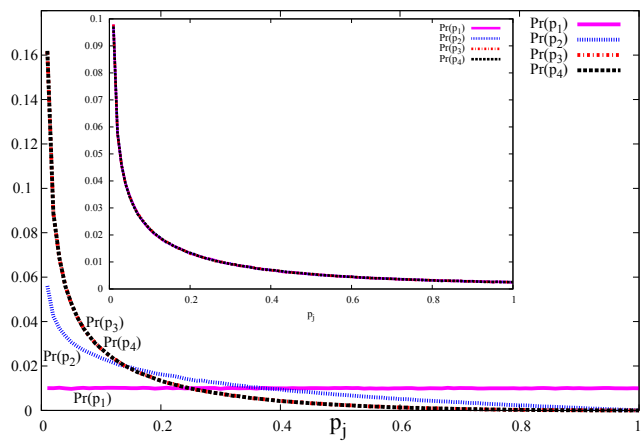}
\protect\caption{(color online) Probability distribution for the components of the biased probability vector $\mathbf{p}=(p_{1},p_{2},p_{3},p_{4})$ for one million random samples of it. In the inset is shown the probability distribution for the components of the unbiased probability vector $\mathbf{q}=(q_{1},q_{2},q_{3},q_{4})$.}
\label{fig_stat_prob_dist}
\end{center}
\end{figure}

\begin{figure}
\begin{center}
\includegraphics[scale=0.6]{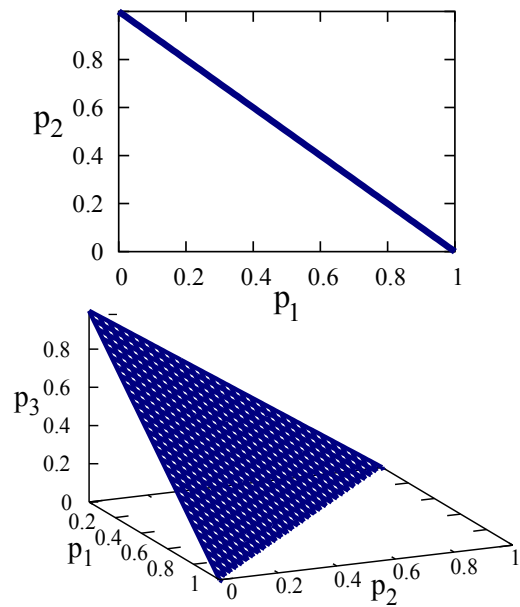}
\protect\caption{(color online) In blue is shown the set of possible values for the
components of the probability vector in the cases of $d=2$ (figure
on the top) and for $d=3$ (figure on the bottom). For higher dimensions
the probability space is a hyperplane.}
\label{fig_prob_space}
\end{center}
\end{figure}

The probability distributions for the four components of the probability vector $\mathbf{p}=(p_{1},p_{2},p_{3},p_{4})$ are shown in Fig. \ref{fig_stat_prob_dist}. For the sake of illustration, the spaces for the pRPV are sketched
geometrically in Fig. \ref{fig_prob_space} for the cases $d=2$ and
$d=3$. We observe that the procedure for generating $\mathbf{p}$ as
explained above is the motivation for the name of the method, the
\emph{normalization method}.

A simple \emph{solution for the biasing problem} just discussed is shuffling
the components of the pRPV in each run of the numerical experiment.
This can be done, for example, by generating a random permutation
of $\{1,2,\cdots,d-1,d\}$, let us call it $\{k_{1},k_{2},\cdots,k_{d-1},k_{d}\}$,
and defining a new pRPV as 
\begin{eqnarray}
\vec{q} & = & (q_{1},q_{2},\cdots , q_{d-1},q_{d}) \nonumber \\
 & := & (p_{k_{1}},p_{k_{2}},\cdots , p_{k_{d-1}},p_{k_{d}}).
\end{eqnarray}
In the table below is presented the mean value of the components of
$\mathbf{q}$ for $10^{6}$ pRPV generated.

\begin{center}
\begin{tabular}{|c|c|c|c|c|c|c|}
\hline 
$d$ & $\langle q_{1}\rangle$ & $\langle q_{2}\rangle$ & $\langle q_{3}\rangle$ & $\langle q_{4}\rangle$ & $\langle q_{5}\rangle$ 
\tabularnewline
\hline 
\hline 
$2$ & $0.5005$ & $0.4995$ &  & \multirow{2}{*}{} & \multirow{3}{*}{} 
\tabularnewline
\cline{1-4} 
$3$ & $0.3332$ & $0.3331$ & $0.3337$ &  &   \tabularnewline
\cline{1-5} 
$4$ & $0.2494$ & $0.2503$ & $0.2499$ & $0.2504$ &   \tabularnewline
\cline{1-6} 
$5$ & $0.2001$ & $0.1997$ & $0.2006$ & $0.1999$ & $0.1997$ 
 \tabularnewline
\hline 
\end{tabular}
\par\end{center}

In the inset of Fig. \ref{fig_stat_prob_dist} is shown an example with the resulting probability distributions for the four components of $\mathbf{q}=(q_{1},q_{2},q_{3},q_{4})$.

From the discussion above we see that in addition to the $d-1$ pRN needed
for the biased pRPV, we have to generate more $d-1$ pRN for the shuffling
used in order to obtain an unbiased pRPV (because $\sum_{j=1}^{d}j=d(d+1)/2$),
resulting in a total of $2(d-1)$ pRN per pRPV.

\section{The trigonometric method}

\label{sec_trigo}

As the name indicates, this method uses a trigonometric parametrization
\cite{Vedral_Prob_trigo,Fritzche4} for the components of the probability
vector $\vec{p}=(p_{1},\cdots,p_{d})$:
\begin{equation}
p_{j}:=\sin^{2}\theta_{j-1} \prod_{k=j}^{d-1}\cos^{2}\theta_{k},\label{eq:trig_param}
\end{equation}
with $\theta_{0}=\pi/2$ (so the name \emph{trigonometric method}).
More explicitly,
\begin{eqnarray}
 &  & p_{1}=\sin^{2}\theta_{0}\cos^{2}\theta_{1}\cos^{2}\theta_{2}\cos^{2}\theta_{3}\cdots\cos^{2}\theta_{d-1}\nonumber \\
 &  & p_{2}=\sin^{2}\theta_{1}\cos^{2}\theta_{2}\cos^{2}\theta_{3}\cdots\cos^{2}\theta_{d-1}\nonumber \\
 &  & p_{3}=\sin^{2}\theta_{2}\cos^{2}\theta_{3}\cos^{2}\theta_{4}\cdots\cos^{2}\theta_{d-1}\nonumber \\
 &  & \vdots\nonumber \\
 &  & p_{d-1}=\sin^{2}\theta_{d-2}\cos^{2}\theta_{d-1}\nonumber \\
 &  & p_{d}=\sin^{2}\theta_{d-1}.
\end{eqnarray}
A simple application of the equality $\cos^{2}\theta_{j}+\sin^{2}\theta_{j}=1$
to this last equation shows that $p_{j}\ge0$ and $\sum_{j=1}^{d}p_{j}=1$.
Therefore this parametrization, which utilizes $d-1$ angles $\theta_{j}$,
leads to a well defined probability distribution $\{p_{j}\}_{j=1}^{d}$.

Let us regard the numerical generation of an unbiased pseudo-random probability
vector by starting with the parametrization in Eq. (\ref{eq:trig_param}).
Of course, this task is accomplished if the components of the pRPV are created with uniform probability distributions. Thus we can proceed as follows. We begin
with $p_{d}$ and go all the way to $p_{1}$ imposing that each $p_{j}$
must be uniformly distributed in $[0,1]$. Thus we must have
\begin{equation}
\theta_{d-1}=\arcsin\sqrt{t_{d-1}}
\end{equation}
and the other angles $\theta_{j}$, with $j=1,\cdots,d-2$, should
be generated as shown in the next equation:
\begin{equation}
\theta_{j}=\arcsin\sqrt{\frac{t_{j}}{\prod_{k=j+1}^{d-1}\cos^{2}\theta_{k}}},\label{eq:thetas}
\end{equation}
where $t_{j}$, with $j=1,\cdots,d-1$, are pseudo-random numbers
with uniform distribution in the interval $[0,1]$. For obvious reasons,
this manner of generating a pRPV is very unstable, and therefore inappropriate,
for numerical implementations. 

A possible way out of this nuisance
is simply to ignore the squared cosines in the denominator of Eq.
(\ref{eq:thetas}). That is to say, we may generate the angles using e.g.
\begin{equation}
\theta_{j}=\arccos\sqrt{t_{j}}
\end{equation}
for all $j=1,\cdots,d-1$. This procedure will give us an uniform
distribution for $\cos^{2}\theta_{j}$ and $\sin^{2}\theta_{j}$,
but will also increase the chance for the components $p_{j}$ with
more terms to have values closer to zero. Thus, there is the issue
of a biased pRPV again. A possible solution for this problem is, once
more, shuffling. In the next two tables are shown the average value
of the components of $10^{6}$ pRPV generated via the trigonometric
method before, $\langle p_{j}\rangle$, and after, $\langle q_{j}\rangle$, shuffling.

\begin{center}
\begin{tabular}{|c|c|c|c|c|c|c|}
\hline 
$d$ & $\langle p_{1}\rangle$ & $\langle p_{2}\rangle$ & $\langle p_{3}\rangle$ & $\langle p_{4}\rangle$ & $\langle p_{5}\rangle$ 
\tabularnewline
\hline 
\hline 
$2$ & $0.4999$ & $0.5001$ &  & \multirow{2}{*}{} & \multirow{3}{*}{} 
\tabularnewline
\cline{1-4} 
$3$ & $0.2502$ & $0.2498$ & $0.4999$ &  &  \tabularnewline
\cline{1-5} 
$4$ & $0.1250$ & $0.1251$ & $0.2497$ & $0.5003$ &   \tabularnewline
\cline{1-6} 
$5$ & $0.0625$ & $0.0624$ & $0.1250$ & $0.2496$ & $0.5006$  \tabularnewline
\hline 
\end{tabular}
\par\end{center}

\begin{center}
\begin{tabular}{|c|c|c|c|c|c|c|}
\hline 
$d$ & $\langle q_{1}\rangle$ & $\langle q_{2}\rangle$ & $\langle q_{3}\rangle$ & $\langle q_{4}\rangle$ & $\langle q_{5}\rangle$
\tabularnewline
\hline 
\hline 
$2$ & $0.5006$ & $0.4994$ &  & \multirow{2}{*}{} & \multirow{3}{*}{} 
\tabularnewline
\cline{1-4} 
$3$ & $0.3331$ & $0.3334$ & $0.3337$ &  &   \tabularnewline
\cline{1-5} 
$4$ & $0.2502$ & $0.2497$ & $0.2501$ & $0.2450$ &   \tabularnewline
\cline{1-6} 
$5$ & $0.2005$ & $0.1998$ & $0.1999$ & $0.2001$ & $0.1997$  \tabularnewline
\hline 
\end{tabular}
\par\end{center}

\begin{figure}[h]
\begin{center}
\includegraphics[scale=0.35]{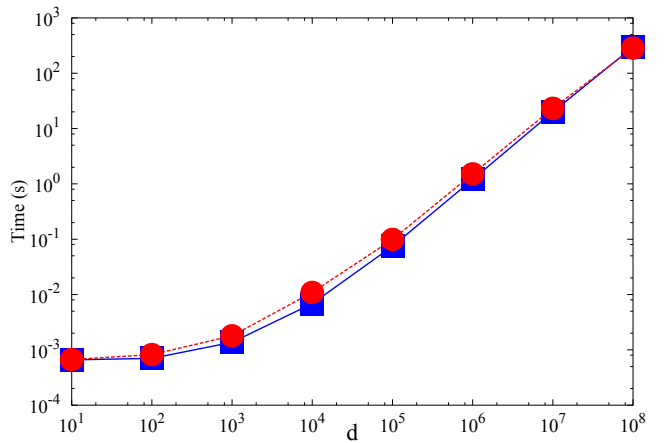}
\protect\caption{(color online) Log-log plot of the computation time required by the
trigonometric method (blue squares) and by the normalization method
(red circles) to generate a pseudo-random probability vector with dimension
equal to $d$. The calculations were realized using a Processor 1.3
GHz Intel Core i5.}
\label{fig_time_d}
\end{center}
\end{figure}

As was the case with the normalization method, for the trigonometric
method we need to generate $2(d-1)$ pRN per pRPV, $d-1$ for the
angles and $d-1$ for the random permutation. Nevertheless, because
of the additional multiplications in Eq. (\ref{eq:trig_param}), the
computation time for the last method is in general a little greater
than that for the former, as is shown in Fig. \ref{fig_time_d}.

\begin{figure}[t]
\centering
\includegraphics[scale=0.365]{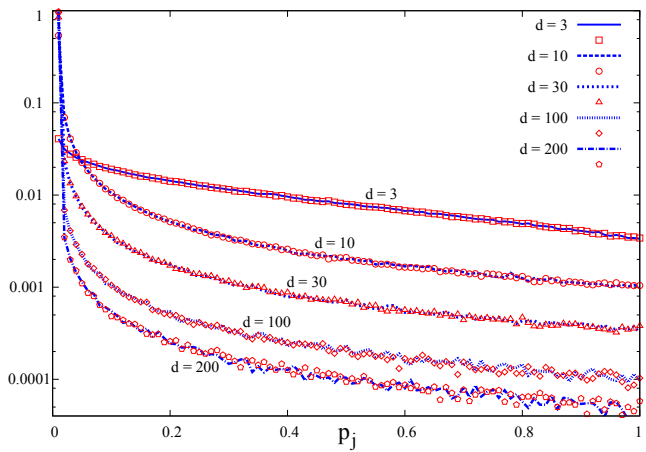}
\protect\caption{(color online) Semi-log plot of the probability distributions for a component of the unbiased probability vectors generated using the trigonometric (lines) and normalization (points) methods for some values of $d$. We see that the two methods yield, for all practical purposes, the same probability distributions for the components of the pRPV.}
\label{fig_comparison}
\end{figure}

\begin{figure}[t]
\centering
\includegraphics[scale=0.35]{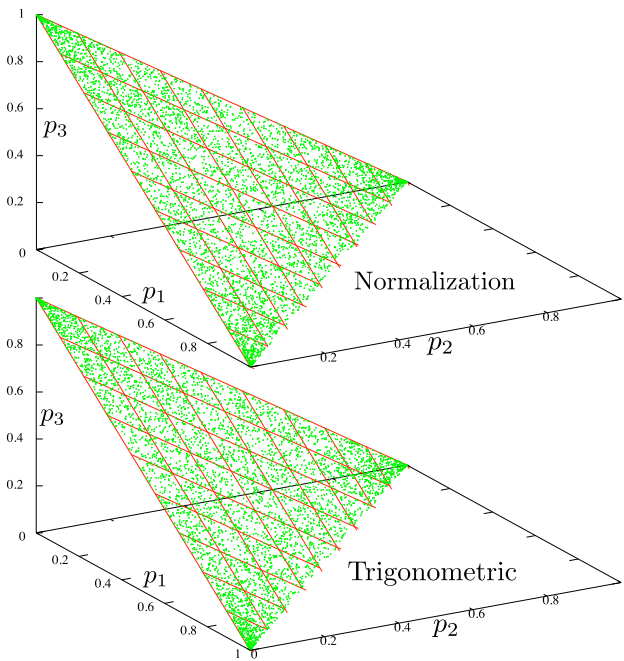}
\protect\caption{(color online) Sample with five thousand pRPV generated using the indicated method. We see that, in this case for which $d=3$, with exception of the slight overpopulated corners, we get a fairly uniform distribution of points in the probability space.}
\label{fig_sample_prob_space}
\end{figure}

One may wonder if the normalization and trigonometric methods, that are at first sight distinct, lead to the same probability distributions for the pRPV's components and also if they produce an uniform distribution for the generated points in the probability hyperplane. We provide some evidences for positive answers for both questions in Figs. \ref{fig_comparison} and \ref{fig_sample_prob_space}, respectively.

\section{Concluding remarks}

\label{conclusions}

In this article we discussed thoroughly three methods for generating
pseudo-random discrete probability distributions. 
We showed that the iid method is not a suitable choice for the problem studied here and identified some
difficulties for the numerical implementation of the trigonometric
method. The fact that in a direct application of both the normalization
and trigonometric methods one shall generate biased probability vectors
was emphasized. Then the shuffling of the pseudo-random probability
vector components was shown to solve this problem at the cost of the
generation of additional $d-1$ pseudo-random numbers for each pRPV. 

It is worthwhile recalling that pure quantum states in $\mathbb{C}^{d}$ can be written in terms of the computational basis $\{|c_{j}\rangle\}_{j=1}^{d}$ as follows:
\begin{equation}
 |\psi\rangle = \textstyle{\sum_{j}}c_{j}|c_{j}\rangle = \sum_{j} |c_{j}|\mathrm{e}^{\phi_{j}}|c_{j}\rangle = \sum_{j} \sqrt{p_{j}}\mathrm{e}^{\phi_{j}}|c_{j}\rangle, 
\end{equation}
where $c_{j}\in\mathbb{C}$ and $\phi_{j}\in\mathbb{R}$. The normalization of $|\psi\rangle$ implies that the set $\{p_{j}\}$ is a probability distribution. Thus the results reported in this article are seem to have a rather direct application for the generation of \emph{unbiased pseudo-random state vectors}. 

We observe however that the content presented in this article can be useful not only for the generation of pseudo-random quantum states in quantum information
science, but also for stochastic numerical simulations in other areas
of science. An interesting problem for future investigations is with
regard to the possibility of decreasing the number of pRN, and thus
the computer time, required for generating an unbiased pRPV. 

\begin{acknowledgements}
This work was supported by the Brazilian funding agencies:
Conselho Nacional de Desenvolvimento Cient\'ifico e Tecnol\'ogico
(CNPq) and Instituto Nacional de Ci\^encia e Tecnologia de Informa\c{c}\~ao
Qu\^antica (INCT-IQ). 
We thank the Group of Quantum Information and Emergent Phenomena and the Group of Condensed Matter Theory at Universidade Federal de Santa Maria for stimulating discussions. We also thank the Referee for his(her) constructive comments.
\end{acknowledgements}


\end{document}